\documentclass[twocolumn,prc,superscriptaddress,unsortedaddress,showpacs,preprintnumbers,amsmath,amssymb]{revtex4}

\usepackage{graphicx}
\usepackage{dcolumn}
\usepackage{bm}

\def\beq{\begin{equation}}
\def\eeq{\end{equation}}
\def\bea{\begin{eqnarray}}
\def\eea{\end{eqnarray}}

\def\fun#1#2{\lower3.6pt\vbox{\baselineskip0pt\lineskip.9pt
  \ialign{$\mathsurround=0pt#1\hfil##\hfil$\crcr#2\crcr\sim\crcr}}}

\begin{document}
\preprint{}

\title{ Proton radioactivity within a generalized liquid drop model}

\author{J. M. Dong}
\affiliation{School of Nuclear Science and Technology, Lanzhou
University, Lanzhou 730000, China}
\author{H. F. Zhang}
\affiliation{School of Nuclear Science and Technology, Lanzhou
University, Lanzhou 730000, China}
\author{G. Royer}
\affiliation{Laboratoire Subatech, UMR: 
IN2P3/CNRS-Universit\'e-Ecole des Mines, Nantes 44, France}

\date{\today}

\begin{abstract}
The proton radioactivity half-lives of spherical proton emitters are
investigated theoretically. The potential barriers preventing the
emission of proton are determined in the quasimolecular shape path
within a generalized liquid drop model (GLDM) including the
proximity effects between nuclei in a neck and the mass and charge
asymmetry. The penetrability is calculated in the WKB
approximation. The spectroscopic factor has been taken into account
in half-life calculation, which is obtained by employing the
relativistic mean field (RMF) theory combined with the BCS method
with the force NL3. The half-lives within the GLDM are compared with
the experimental data and other theoretical values. GLDM
works quite well for spherical proton emitters when the
spectroscopic factors are considered, indicating the necessity of
introducing the spectroscopic factor and the success of the GLDM for
proton emission. Finally, we present two formulae for proton
emission half-life similar to the Viola-Seaborg formulae and
Royer's formulae of $\alpha$-decay.
\end{abstract}

\pacs{23.50.+z, 21.10.Tg, 21.10.Jx, 21.60.-n}

\maketitle

\section{Introduction}\label{intro}\noindent

The opportunities provided by the radioactive beam facilities make
the study of the exotic nuclei with extreme numbers of neutrons or
protons a very interesting topic both from the experimental and
theoretical points of view. And studies on these exotic nuclei lead
to the discovery of a new form of radioactivity-proton emission. The
proton drip line represents one of the fundamental limits of nuclear
existence and the nucleus with a very large excess of protons
undergo spontaneous proton emission towards stability. Besides, the
rapid proton capture process which plays a very important role in
nuclear astrophysics has its inverse in the proton radioactivity
from the nuclear ground state or low isomeric states. Therefore the
study of the proton emission is significative. This proton emission
from nuclei was firstly observed in an isomeric state of $^{53}$Co
in 1970. With the development of experimental facilities and
radioactive beams, proton emissions from ground state or low
isomeric states have been identified between $Z=51$ and $Z=83$
\cite{AA} and more proton-emitting nuclei will be observed in
experiments in the future.

The proton radioactivity can be used as an useful tool to extract
nuclear structure information such as the shell structure and the
coupling between bound and unbound nuclear states \cite{shell}.
Measurement on the proton energy, half-life and proton branching
ratio (fine structure) helps to determine the angular momentum $l$
carried away by the emitted proton and to characterize its wave
function inside the nucleus \cite{AA,AT,LS}. The proton emission can
be dealt with in the framework of WKB barrier penetration model
since the decay process can be treated in a simple quantum tunneling
effect through a potential barrier. Several approaches have been
employed to study the half-lives of spherical proton emitters, such
as the distorted-wave Born approximation \cite{Bonn}, the
density-dependent M3Y(DDM3Y) effective interaction \cite{DDM3Y,MBG},
JLM interaction \cite{MBG}, and the unified fission model \cite{UF}.
However, the calculations in Ref. \cite{DDM3Y,MBG,UF} did not take
account the spectroscopic factor. The spectroscopic factor is
very important from the viewpoint of the nuclear structure. In this
study, we calculate the spectroscopic factor by employing the
relativistic mean field theory in combination with the BCS method,
and then determine the partial proton emission half-lives of
spherical proton emitters within the macro-microscopic GLDM. The
calculated results are compared with the experimental data and other
theoretical results.

The paper is organized as follows. In Sec. II, the framework of GLDM
and the method for spectroscopic factor are presented. The
calculated proton emission half-lives are shown and discussed in
Sec. III. In Sec. IV, two formulae for proton emission are proposed.
Finally, we present a brief summary of the present work.

\section{methods}\label{GL}\noindent

The GLDM allows to describe the processes of fusion, fission, light
nucleus and $\alpha$ emission
\cite{fis,fus,GLDM,RG4,RG2,RG3,Zh1,GRHF,DONG0}. The macroscopic
energy includes the volume, surface, Coulomb and proximity energies:
\begin{equation}
E=E_{V}+E_{S}+E_{C}+E_{\text{Prox}}.
\end{equation}
For one-body shapes, the volume, surface and Coulomb energies are
given by:
\begin{equation}
E_{V}=-15.494(1-1.8I^{2})A \ \textrm{MeV},
\end{equation}
\begin{equation}
E_{S}=17.9439(1-2.6I^2)A^{2/3}(S/4\pi R_0^2) \ \textrm{MeV},
\end{equation}
\begin{equation}
E_{C}=0.6e^2(Z^2/R_0) \times 0.5\int
(V(\theta)/V_0)(R(\theta)/R_0)^3 \sin \theta d \theta.
\end{equation}
$I$ is the relative neutron excess and $S$ is the surface area of
the one-body deformed nucleus. $V(\theta )$ is the electrostatic
potential at the surface and $V_0$ the surface potential of the
sphere. When the nuclei are separated :
\begin{equation}
E_{V}=-15.494\left \lbrack (1-1.8I_1^2)A_1+(1-1.8I_2^2)A_2\right
\rbrack \ \textrm{MeV},
\end{equation}
\begin{equation}
E_{S}=17.9439\left
\lbrack(1-2.6I_1^2)A_1^{2/3}+(1-2.6I_2^2)A_2^{2/3} \right \rbrack \
\textrm{MeV},
\end{equation}
\begin{equation}
E_{C}=0.6e^2Z_1^2/R_1+0.6e^2Z_2^2/R_2+e^2Z_1Z_2/r.
\end{equation}
The additional centrifugal energy $E_{l}$ coming from the angular
momentum of the emitted proton has been introduced:
\begin{equation}
E_{l}(r)=\frac{\hbar^{2}}{2\mu }\frac{l(l+1)}{r^{2}}.
\end{equation}
Here $A_i$, $Z_i$, $R_i$ and $I_i$ are the mass number, charge
number, radii and relative neutron excesses of the two nuclei. The
relative neutron excess of a proton is fixed at $I_{2}=0$ to ensure
a negative volume energy and a positive surface energy. The
dimensionless quantity of $l$ is the angular momentum carried by the
emitted proton(angular momentum transfer). $\mu$ is the reduced
mass. $r$ is the distance between the mass centers. The radii
$R_{i}$ of the daughter nucleus and proton are given by \cite{RG2}:
\begin{equation}
R_i=(1.28A_i^{1/3}-0.76+0.8A_i^{-1/3}) \ \textrm{fm}, i=1,2.
\end{equation}
The radii $R_{0}$ of parent nucleus can be obtained with the volume
conservation:
\begin{equation}
R_{0}^{3}=R_{1}^{3}+R_{2}^{3}.
\end{equation}

The surface energy results from the effects of the surface tension
forces in a half space. The nuclear proximity energy
$E_{\text{prox}}$ has been introduced to take into account the
additional surface effects due to the attractive nuclear forces
between the surfaces in a neck or a gap between two separated
fragments:
\begin{equation}
E_{\text{Prox}}(r)=2\gamma \int _{h_{\text{min}}} ^{h_{\text{max}}}
\Phi \left \lbrack D(r,h)/b\right \rbrack 2 \pi hdh,
\end{equation}
where $h$ is the distance varying from the neck radius or zero to
the height of the neck border. $D$ is the distance between the
surfaces in regard and $b=0.99$ fm is the surface width. $\Phi$ is
the proximity function of Feldmeier \cite{R3} and the surface
parameter $\gamma$ is the geometric mean between the surface
parameters of the two nuclei or fragments:
\begin{equation}
\gamma =0.9517\sqrt{(1-2.6I_{1}^{2})(1-2.6I_{2}^{2})}\text{ \
MeV}\cdot \text{fm}^{-2}.
\end{equation}

The partial half-life of a spherical proton emitter is calculated
using the WKB barrier penetration probability. The decay constant of
the proton emitter is simply defined as
$\lambda=\nu_{0}PS_{\text{p}}$ and half-life
$T_{\text{p}}=\textrm{ln}2/{\lambda}$. The assault frequency
$\nu_{0}$ has been taken as $8\times10^{20}$ s$^{-1}$. The barrier
penetrability $P$ is calculated by the following formula:
\begin{equation}
P=\exp \left[ -\frac{2}{\hbar }\int_{R_{\text{in}}}^{R_{\text{out}}}\sqrt{%
2B(r)(E(r)-E(\text{sphere}))}dr\right].
\end{equation}
where $R_{\text{in}}$ and $R_{\text{out}}$ are the two turning
points of the WKB action integral. The two following approximations
are used here: $R_{\text{in}}=R_{1}+R_{2}$ and $B(r)=\mu$.
$R_{\text{out}}$ is given as:
\begin{equation}
R_{\text{out}}=\frac{Z_{1}Z_{2}e^{2}}{2Q}+\sqrt{\left( \frac{Z_{1}Z_{2}e^{2}%
}{2Q}\right) ^{2}+\frac{l(l+1)\hbar ^{2}}{2\mu Q}}.
\end{equation}

For proton radioactivity, the spectroscopic factor is given by
\cite{Bonn,DSD}:
\begin{equation}
S_{\text{p}}=u_{j}^{2},
\end{equation}
where $u_{j}^{2}$ is the probability that the spherical orbit of
emitted proton is empty in the daughter nucleus. Fortunately, the
daughter nuclei of spherical proton emitters are all in ground
states. Thus, it is relatively easy to determine this
spectroscopic factor by using the RMF theory combined with the BCS
method. The RMF automatically includes the spin-orbit interaction.
It has received much attention due to its great success in describing
the structure of the stable nuclei~\cite{GG}, neutron-rich
nuclei~\cite{JM1}, proton-rich nuclei~\cite{GALS}, superdeformed
nuclei~\cite{JKPR} and superheavy nuclei~\cite{ZZR,SDG,HFZ2} . It
is now a standard tool in low energy nuclear structure. The RMF
theory is well known and it will not be discussed in detail here.
The pairing correlation is treated by the BCS method. We have
introduced the strength of the pairing forces in the following
forms for neutrons and protons, respectively~\cite{HFZ2}:
\begin{eqnarray}
G_{\text{n}} &=&\frac{21}{A}(1-\frac{N-Z}{2A})\text{ \ MeV ,} \\
\text{\ }G_{\text{p}} &=&\frac{27}{A}(1+\frac{N-Z}{2A})\text{ \
MeV\,\ }
\end{eqnarray}
which depend on the proton number $Z$  and neutron number $N$. $A$
is the total mass number. The NL3 parameter set, which has been used
with enormous success in the description of a variety of
ground-state properties of spherical, deformed and exotic
nuclei~\cite{GLJK,BGT}, is used here. Unlike the situation near the
neutron drip line, for proton-rich nuclei the Coulomb barrier
confines the protons in the interior of the nucleus. As a
consequence, the effects of the coupling to the continuum is weaker
and therefore for nuclei close to proton drip line, the RMF+BCS
model could still be considered as a reasonable approximation
providing sufficiently accurate results~\cite{GALS}.

\section{Half-lives of spherical proton emitters }\label{HF}\noindent

\begin{table*}[llllllllllllllll]
\label{table1} \caption{Comparison between experimental and
calculated proton radioactivity logarithmic half-lives of spherical
proton emitters. The asterisk(*) symbols in parent nuclei denote the
isomeric states. The experimental data of $^{155}$Ta and $^{159}$Re
are taken from Ref. \cite{RDP} and Ref. \cite{DTJ}, respectively and
their $Q$ values are calculated using the measured emitted proton
energies. Other experimental data are from Ref. \cite{AA}.}
\begin{ruledtabular}
\begin{tabular}{llllllllllllllll}
Parent& \emph{l} & $Q$(MeV) & Penetrability & $S_{\text{p}}$ &$\log _{10}T_{\text{p}}(\text{s})$ & $\log _{10}T_{\text{p}}(\text{s})$ &$\log _{10}T_{\text{p}}(\text{s})$ &$\log _{10}T_{\text{p}}(\text{s})$ &$\log _{10}T_{\text{p}}(\text{s})$    \\
& & expt. & & & expt. & GLDM & DDM3Y\cite{DDM3Y}& DDM3Y\cite{MBG}&JLM\cite{MBG} \\
\hline

$^{105}$Sb & 2 & 0.491&$1.280\times10^{-23}$&0.999  &2.049  &1.831 &1.97&  2.27&1.69\\
$^{145}$Tm &5 & 1.753& $6.759\times10^{-16}$&0.580 &-5.409  &-5.656 &-5.14&-5.20&-5.10\\
$^{147}$Tm& 5 & 1.071&$3.931\times10^{-22}$&0.581  &0.591  &0.572 &0.98&0.98&1.07\\
$^{147}$Tm*& 2 & 1.139&$2.504\times10^{-18}$&0.953  &-3.444  &-3.440 &-3.39&-3.26&-3.27\\
$^{150}$Lu& 5 & 1.283&$3.554\times10^{-20}$&0.497  &-1.180  &-1.309 &-0.58&-0.59&-0.49\\
$^{150}$Lu*& 2 & 1.317&$5.734\times10^{-17}$&0.859  &-4.523  &-4.755 &-4.38&-4.24&-4.24\\
$^{151}$Lu& 5 & 1.255&$1.839\times10^{-20}$&0.490  &-0.896  &-1.017 &-0.67&-0.65&-0.55\\
$^{151}$Lu*& 2 & 1.332&$8.262\times10^{-17}$&0.858  &-4.796  &-4.913 &-4.88&-4.72&-4.73\\
$^{155}$Ta& 5 & 1.453&$5.280\times10^{-19}$&0.422  &-2.538  &-2.410 &-4.65&-4.67&-4.57\\
$^{156}$Ta& 2 & 1.028&$4.994\times10^{-21}$&0.761  &-0.620  &-0.642 &-0.38&-0.22&-0.23\\
$^{156}$Ta*& 5 & 1.130 &$1.793\times10^{-22}$&0.493 &0.949  &0.991 &1.66&1.66&1.76\\
$^{157}$Ta& 0 & 0.947&$1.608\times10^{-21}$&0.797  &-0.523  &-0.170 &-0.43&-0.21&-0.23\\
$^{159}$Re& 5 & 1.816 &$1.216\times10^{-16}$&0.308 &-4.678  &-4.636 &--&--&--\\
$^{160}$Re& 2 & 1.284&$2.204\times10^{-18}$&0.507  &-3.046  &-3.111 &-3.00&-2.86&-2.87\\
$^{161}$Re& 0 & 1.214&$2.024\times10^{-18}$&0.892  &-3.432  &-3.319 &-3.46&-3.28&-3.29\\
$^{161}$Re*& 5 & 1.338&$1.419\times10^{-20}$&0.290  &-0.488  &-0.677 &-0.60&-0.57&-0.49\\
$^{164}$Ir& 5 & 1.844 &$7.542\times10^{-17}$&0.188 &-3.959  &-4.214 &-3.92&-3.95&-3.86\\
$^{165}$Ir*& 5 & 1.733&$1.335\times10^{-17}$&0.187  &-3.469  &-3.460 &-3.51&-3.52&-3.44\\
$^{166}$Ir& 2 & 1.168 &$2.624\times10^{-20}$&0.415 &-0.824  &-1.099 &-1.11&-0.96&-0.96\\
$^{166}$Ir*& 5 & 1.340&$4.887\times10^{-21}$&0.188  &-0.076  &-0.025 &0.21&0.22&0.30\\
$^{167}$Ir& 0 & 1.086 &$1.126\times10^{-20}$&0.912 &-0.959  &-1.074 &-1.27&-1.05&-1.07\\
$^{167}$Ir*& 5 & 1.261 &$6.559\times10^{-22}$&0.183 &0.875  &0.858 &0.69&0.74&0.81\\
$^{171}$Au& 0 & 1.469 &$7.608\times10^{-17}$&0.848 &-4.770  &-4.872 &-5.02&-4.84&-4.86\\
$^{171}$Au*& 5 & 1.718 &$4.101\times10^{-18}$&0.087 &-2.654  &-2.613 &-3.03&-3.03&-2.96\\
$^{177}$Tl& 0 & 1.180 &$1.324\times10^{-20}$&0.733 &-1.174  &-1.049 &-1.36&-1.17&-1.20\\
$^{177}$Tl*& 5 & 1.986&$1.166\times10^{-16}$&0.022  &-3.347  &-3.471 &-4.49&-4.52&-4.46\\
$^{185}$Bi& 0 & 1.624 &$1.942\times10^{-16}$&0.011 &-4.229  &-3.392 &-5.44&-5.33&-5.36\\
\end{tabular}
\end{ruledtabular}
\end{table*}

The values of angular momentum transfer $l$, decay energy $Q$,
penetrability $P$, spectroscopic factor $S_{\text{p}}$, the
experimental and calculated proton emission half-life are given in
Table I. The experimental $Q$ value, which is a crucial quantity to
determine the decay half-life, is used for the calculation and the
potential barrier given by the GLDM has been adjusted to reproduce
the experimental $Q$ value.

In Ref. \cite{DDM3Y}, D. N. Basu and coworkers calculated the
half-lives of spherical proton emitters using the density-dependent
M3Y effective interaction with the density of the daughter nucleus
taking from phenomenological models. Madhubrate Bhattacharya and G.
Gangopadhyay developed this model by obtaining this density from
mean field calculation and also used the JLM effective interaction
\cite{MBG}. The results have been presented in the last three
columns of Table I for comparison. As can be seen, The DDM3Y and JLM
models can provide good explanation for most cases. Their is no
doubt that the DDM3Y and JLM models are very successful because
their microscopic nature includes many nuclear features. In a number
of decays, however, the results do not match so well, such as
$^{147}$Tm, $^{150}$Lu, $^{156}$Ta and $^{156}$Ta(isomeric state).
For $^{177}$Tl(isomeric state) and $^{185}$Bi, the discrepancies are
off by an order of magnitude compared with the experimental data. It
is their calculations without introducing the spectroscopic factor
that leads to a large deviation for some nuclei. In order to obtain
more information about proton-radioactivity and preform more
accurate investigation theoretically, we calculated the
spectroscopic factor within the RMF+BCS, and the results are shown
in the fifth column of Table I. One could notice that, spectroscopic
factor is $S_{\text{p}}\sim 1$ at the beginning of a proton shell of
residual daughter nucleus, but moving away, it decreases, becoming
quite small at the end of the shell. This shows the shell structure
plays an important role for the spectroscopic factor. Hence the
spectroscopic factor includes the shell effect to a large extent.
For $^{177}$Tl (isomeric state) and $^{185}$Bi, the spectroscopic
factors are very small, leading to the longer half-lives than the
DDM3Y and JLM calculated.

We calculate the half-lives by employing the GLDM taking account the
spectroscopic factors. The penetration probabilities obtained with
GLDM are shown in the fourth column. The penetrability $P$ stay
between $10^{-23}$ and $10^{-16}$ which are relatively very large
while the range is narrow, compared with $10^{-39}\sim 10^{-14}$ for
$\alpha$-decay \cite{HFR}. Hence it is easy for the proton to escape
from the proton emitter, confirming that the proton-emitting nuclei
are weakly bound. The calculated half-lives are presented in the
seventh column. The half-lives by the GLDM with the average
discrepancy less than $40\%$, give better agreement with the
experimental data than that by DDM3Y and JLM models. For nuclei
$^{147}$Tm, $^{150}$Lu, $^{156}$Ta, $^{156}$Ta(isomeric state) and
$^{177}$Tl(isomeric state), which the DDM3Y and JLM can not explain
well, the GLDM could give a very excellent results with the
deviation $4\%$, $26\%$, $5\%$, $10\%$ and $33\%$ respectively. This
indicates that the including of the spectroscopic factors in
calculations is necessary. The quantitative agreement with the
experimental data are better than other theoretical ones which 
demonstrates that the GLDM with the proximity effects, centrifugal
potential energy, the mass asymmetry and spectroscopic factor could
be used to investigate the proton emission successfully when the
right $Q$ values are given. The GLDM overestimates the half-life of
$^{185}$Bi by one order of magnitude. The first reason possibly is
the uncertainty of the $Q$ value or the emitted proton being not in
s$_{1/2}$ state, and it requires further investigation theoretically
and measurements with high accuracy. The second reason is, perhaps,
the shell effect is not included in the potential barrier, though
the most shell effect contribution has been included in the $Q$
value and spectroscopic factor. It is sure that the GLDM connected
with WKB approximation will quantitatively give more consistent
results for the proton radioactivity half-life when the shell effect
is included in the penetration barrier.

Recently, the spherical proton emitter $^{155}$Ta was observed. Its
emitted proton energy and half-life have been measured again
\cite{RDP}. With the angular momentum transfer $l=5$ \cite{BB} and
proton energy $E_{\text{p}}=1.444\pm0.015$ MeV ($Q=1.453\pm0.015$
MeV), we obtain half-life of $T_{\text{p}}=3.9^{+1.4}_{-1.0}$ ms
using the GLDM compared with the experimental data of
$T_{\text{p}}=2.9^{+1.5}_{-1.1}$ ms. The $Q$ value is compatible
with the half-life, indicating the experimental data should be
reliable. The new spherical proton emitter $^{159}$Re was
synthesized in the reaction $^{106}$Cd ($^{58}$Ni, p4n) $^{159}$Re
\cite{DTJ} and its proton emission $Q$ value along with half-life
have been measured recently. The $Q$ value is compatible with the
half-life if and only if $l=5$ (the calculated value is
$23^{+8}_{-6}$ $\mu$s in contrast with the experimental data of
$21^{+4}_{-4}$ $\mu$s), which indicates the proton is emitted from
an $\pi$h$_{11/2}$ state agreeing with the conclusion in Ref.
\cite{DTJ}.

\section{New formulas for proton emission half-life }\label{HF}\noindent

The centrifugal potential energy $E_{l}$ can reduce the tunneling
probability and hence increases the half-life. A formula can be
deduced to describe this relationship between half-life and $l$
value for proton emission, being similar to the formula for
$\alpha$-decay in Ref. \cite{DDDD}:
\begin{equation}
\log _{10}T_{\text{p}}(l)=\log
_{10}T_{\text{p}}(0)+c_{0}\frac{l(l+1)}{\sqrt{(A-1)(Z-1)A^{-2/3}}}.\label{A}
\end{equation}
The half-life $T_{\text{p}}$ is measured in second. $c_{0}$ is
slightly model dependent with $c_{0}=2.5$ for the GLDM. With the
formula (\ref{A}), we fit new formulae that could be used to
describe half-life for proton emission. A formula is proposed in the
following form:
\begin{eqnarray}
\log_{10}\left[ T_{\text{p}}(\text{s})\right] =(aZ+b)Q^{-1/2}+c+\nonumber\\
           c_{0}\frac{l(l+1)}{\sqrt{(A-1)(Z-1)A^{-2/3}}},\label{BB}
\end{eqnarray}
where $Z$ and $A$ are charge and mass numbers of the parent nucleus
respectively, and $Q$ the proton decay energy in MeV. The first two
terms are similar to Viola-Seaborg formulae \cite{VSS1,VSS2} for
$\alpha$-decay and the last term is exactly the contribution of
centrifugal barrier from formula (\ref{A}). Performing a least
squares fit to the half-lives of first 25 spherical proton emitters
available in Table I, we obtain a set of parameters for formula
(\ref{BB}). There values are: $a=0.3437$, $b=4.9628$, $c=-31.1253$
and $c_{0}=2.5950$, with the average deviation $\overline{\sigma
}$=0.153 between the experimental and formula. Additionally, we
obtained a set of parameters for half-lives of deformed proton
emitters by employing a least squares fit to data, which include 11
nuclei ($Z=53-67$) that could be found in literature \cite{AA} or
\cite{BB}. These parameters are: $a=0.3637$, $b=4.6467$,
$c=-30.9299$ and $c_{0}=2.6244$. The average deviation is
$\overline{\sigma }=0.323$. The half-life increases by 3$\sim $4
orders of magnitude when the angular momentum transfer $l$ is
changed from zero to five in terms of this formula. In other words,
the half-life of proton emission is quite sensitive to the angular
momentum $l$ of the emitted proton, which in turn helps to determine
the $l$ value when half-life and $Q$ value are measured. On the
other hand, so many proton emitters have been observed in
experiments at present due to the centrifugal barriers prolonging
lifetimes of these nuclei to a great extent.

\begin{figure}[htbp]
\begin{center}
\includegraphics[width=0.5\textwidth]{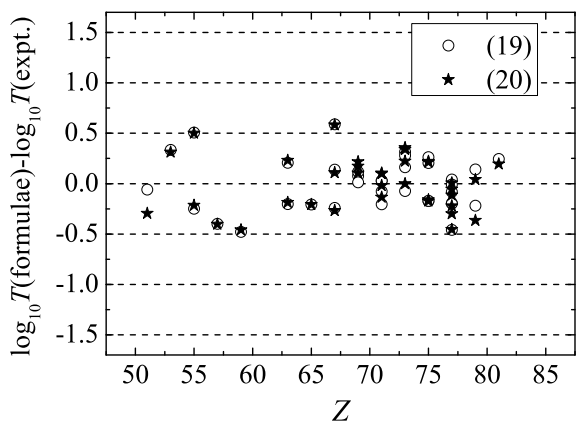}
\caption{Deviation between the formulae (\ref{BB},\ref{D}) and
experimental logarithm of half-lives for proton emission.}
\end{center}
\end{figure}

The other formula for proton emission is given by
\begin{eqnarray}
\log _{10}\left[ T_{\text{p}}(\text{s})\right]
&=&a+bA^{1/6}\sqrt{Z}+cZQ^{-1/2}+  \notag
\\
&&c_{0}\frac{l(l+1)}{\sqrt{(A-1)(Z-1)A^{-2/3}}}.  \label{D}
\end{eqnarray}
The first three terms are similar to Royer's formulae for
$\alpha$-decay \cite{R1,GRHF}. With the same method we discussed
above, we obtained the parameter sets: $a=-23.0632$, $b=-0.4225$,
$c=0.4170$ and $c_{0}=2.5989$ for spherical proton emitters with
average deviation $\overline{\sigma } =0.183$; $a=-23.9341$,
$b=-0.3936$, $c=0.4385$ and $c_{0}=2.6167$ for deformed proton
emitters with $\overline{\sigma } =0.316$. The two formulae
(\ref{BB},\ref{D}) can validate each other, and get the more
reliable results in future study. The experimental proton
radioactivity half-lives of most nuclei can be reproduced within a
factor of less than 2 by the above two formulae. The little
discrepancy suggests that these two formulae could be used to
determine the $l$ value when $Q$ and $T_{\text{p}}$ values have been
measured and then extract some useful information about nuclear
structure as well as to calculate proton emission half-lives. The
proton and $\alpha$ emission can be described by the similar
formulae with the different parameters, which is exactly what we
expect. Since both the two decay modes are quantum tunneling effect,
the studies on them should be unified.

From these two formula, it is easy to deduce the following equation:
\begin{equation}
\frac{\partial \left( \log _{10}T_{\text{p}}\right) }{\partial Q}=-\frac{1}{2%
}(aZ+b)Q^{-3/2},\text{ or }-\frac{1}{2}cZQ^{-3/2},
\end{equation}
which reflects the $Q$ value dependence of half-life. By comparing
these parameters and $Q$ values with that in Viola-Seaborg and
Royer's formulae for $\alpha$-decay, one could find that the
half-life is more sensitive to $Q$ value for proton emission. In
addition, $C_{0}\approx2.6$ for proton emission in contrast with
$C_{0}=1.0$ \cite{DDDD} for $\alpha$-decay indicates that the centrifugal
barrier is much more important for proton emission than that for
$\alpha$-decay, due to smaller reduced mass $\mu$ (and hence the
high centrifugal barrier) compared with that in $\alpha$-decay
system. These imply that it is quite difficult to predict the
half-life of proton emission for unknown nucleus since the $Q$ value
can not be obtained with a good accuracy and since the uncertainty on the
$l$ value is large.

\section{Summary}\label{SUM}\noindent

The proton radioactivity of spherical proton emitters has been
analyzed in the framework of the GLDM for the first time. The
penetration barriers are constructed in the quasimolecular shape
path, and the penetration probabilities are calculated with the
WKB approximation. The penetration probabilities are relatively
very large while its range is narrow. Therefore it is easy for the
proton to escape from the proton emitter, confirming the proton
emitting nuclei are weakly bound. The spectroscopic factors have
been taken into account in half-lives calculations, which are
obtained by employing the relativistic mean field theory combined
with the BCS method. The spectroscopic factor is affected greatly
by the proton shell structure, and in turn it contains shell
effect to a large extent. For $^{177}$Tl (isomeric state) and
$^{185}$Bi, the spectroscopic factors are very small (0.022 and
0.011), leading to the longer half-lives than the DDM3Y and JLM
calculated. Although the DDM3Y and JLM models include the
appropriate considerations in the microscopic level, the present
calculations are better agreement with the experimental data than
that within the DDM3Y and JLM models. This, indicates the
considering of spectroscopic factor is necessary for proton
emission, especially for the nuclei with residual daughter nuclei
at the end of the shell. Additionally, some newly observed proton
emitters $^{155}$Ta and $^{159}$Re have been analyzed. Finally,
two formulae similar to Viola-Seaborg formulae and Royer's
formulae have been proposed for proton radioactivity. On the one
hand, they can be employed to calculate the proton emission
half-life. On the other hand, they can be used to determine the $l$
value when the $Q$ value and half-life have been measured and then
extract some useful information about nuclear structure since the
decay rate is quite sensitive to $l$ value. The proton and
$\alpha$ emission can be described by the similar formulae with
different parameters, but the half-life of proton radioactivity is
more sensitive to $Q$ and $l$ values compared with $\alpha$-decay
according to our analysis, leading to the prediction of the
half-lives for proton emission quite difficult.

\section*{Acknowledgements}

This work is supported by the Natural Science Foundation of China
(grants 10775061, and 10805016); by the Fundamental Research Fund
for Physics and Mathematics of Lanzhou University( grants
LZULL200805); by the CAS Knowledge Innovation Project
NO.KJCX-SYW-N02; The Major State Basic Research Developing Program
of China (2007CB815004).

\end{document}